# Ab-initio simulation and experimental validation of beta-titanium alloys

D. Raabe [a], B. Sander [b], M. Friák [c], D. Ma [d], J. Neugebauer [e]

Max-Planck-Institut für Eisenforschung, Max-Planck-Str. 1, 40237 Düsseldorf, Germany

[a] d.raabe@mpie.de,  [b] b.sander@mpie.de,  [c] m.friak@mpie.de,
[d] d.ma@mpie.de,  [e] j.neugebauer@mpie.de

## Abstract

In this progress report we present a new approach to the ab-initio guided bottom up design of β-Ti alloys for biomedical applications using a quantum mechanical simulation method in conjunction with experiments. Parameter-free density functional theory calculations are used to provide theoretical guidance in selecting and optimizing Ti-based alloys with respect to three constraints: (i) the use of non-toxic alloy elements; (ii) the stabilization of the body centered cubic β phase at room temperature; (iii) the reduction of the elastic stiffness compared to existing Ti-based alloys.

Following the theoretical predictions, the alloys of interest are cast and characterized with respect to their crystallographic structure, microstructure, texture, and elastic stiffness. Due to the complexity of the ab initio calculations, the simulations have been focused on a set of binary systems of Ti with two different high melting bcc metals, namely, Nb and Mo. Various levels of model approximations to describe mechanical and thermodynamic properties are tested and critically evaluated. The experiments are conducted both, on some of the binary alloys and on two more complex engineering alloy variants, namely, Ti-35wt.%Nb-7wt.%Zr-5wt.%Ta and a Ti-20wt.%Mo-7wt.%Zr-5wt.%Ta.

# 1 Introduction

## 1.1 Motivation for the design of beta-titanium alloys as biomaterials

The design of novel Ti-based alloys for biomedical load-bearing implant applications such as hip or knee prostheses aims at providing structural materials which are characterized by a good corrosion stability in the human body, high fatigue resistance, high strength-to-weight ratio, good ductility, low elastic modulus, excellent wear resistance, low cytotoxicity, and a negligible tendency to provoke allergic reactions [1-12].

Commonly, Ti and Ti-base alloys occur in one or a mixture of two basic crystalline structures: the α phase and the β phase which assume hexagonal (hex) and body-centered cubic (bcc) structure, respectively. Thus, three general classes of Ti-base alloys are defined: α, α – β, and β [1-30]. The transition temperature from the α to the β phase is about 882°C for pure Ti. Elements which promote higher or lower transformation temperatures are referred to as α stabilizers (like O, Al, La) or β stabilizers (like Mo, Nb, Ta), respectively.

Recent studies have revealed that a compromise along the biomedical constraints mentioned above can be obtained by designing Ti alloys which use the most bio-compatible elements, i.e. Ta, Mo, Nb, and Zr as alloy ingredients for stabilizing the bcc β phase [6-26]. Experimental investigations on such alloy variants have shown that these materials can indeed match most of the desired properties outlined above, i.e. they do not reveal any toxicity versus osteoblastic cells, have a high corrosion resistance, and show good mechanical properties owing to solid solution and particle strengthening while preserving the light weight character of Ti.

An attractive feature of β Ti alloys is their low elastic modulus which is below that of conventional α or α-β Ti alloys commonly used for biomedical applications. This latter property is of particular relevance because an important requirement of bone-replacing implants is a low elastic stiffness approximating as far as possible that of the surrounding bone tissue [13-25]. The background of this aspect is the stress shielding effect: Bone is a living material which is subject to mechanical loads resulting from the weight it carries and the motion it creates. When an elastically much stiffer surgical implant part replaces a portion of the human skeleton, it takes over a considerable part of that mechanical load, thereby shielding the remaining bone that surrounds and abuts the implant. Reducing the physiological loads on the bone entails resorption mechanisms which lead to a drop in bone density, mineralization state, strength, and health. The mechanical system bone-implant undergoes three main effects upon stress shielding: First, osteoporosis associated with under-utilization that characterizes tissue resorption subsequent to the absence or the decrease in local physiological mechanical stress. Second, tissue resorption increases the danger of formation and migration of wear debris at the bone-implant interface via biological fluid transport. Third, the stiffness gap between bone and material gives rise to micro-motions at the bone-implant interface. Owing to this context, the stress shielding and the described effects associated with it create multiple undesired phenomena at the bone-implant system and may finally lead to contact loosening, premature implant failure, or debris-induced infections [27-30].

This phenomenon explains that in the design of Ti alloys for biomedical applications the reduction of the elastic stiffness must be regarded as a high priority constraint in the overall strategy [6-26]. The major challenge associated with this point is documented by the fact that commercial pure Ti and α–β type Ti-6 wt.%Al-4 wt.%V alloys which are currently widely used as structural biomaterials for the replacement of hard tissues in artificial joints have an elastic modulus of 105-110 GPa while the human cortical bone has a stiffness of about 20 GPa.

New alloys which are dominated by a large amount of stable β phase reveal a significant drop in the elastic modulus to values as small as 65 GPa [16-24]. It is, therefore, likely that the next generation of structural materials for replacing hard human tissue will be of this type. Another reason to develop a next generation of alloys is that conventional two phase α-β variants such as Ti-6wt.%Al-4 wt.%V contain both vanadium, being a cytotoxic element and aluminum, being a suspicious element for causing neurological disorders.

**1.2   Motivation for applying ab initio calculations to the design of beta-titanium alloys**

Owing to the arguments outlined above the aim of the present study is to understand and design biocompatible β-phase Ti alloys with a small elastic modulus. However, in contrast to earlier approaches which used phenomenological rules in conjunction with metallurgical experience for estimating a beneficial alloy composition we pursue in this work a theory-guided bottom up design strategy to achieve that goal. The approach is based on the combination of ab-initio simulations with experiments. The ab-initio predictions are based on density functional theory (DFT) [31,32] in the generalized gradient approximation (GGA-PBE96) [33]. They are used to provide theoretical guidance in selecting suited Ti-based alloys with respect to the three constraints of first, using exclusively non-toxic alloy elements; second, obtaining a stable body centered cubic β phase; and third, reducing the elastic stiffness.

Following the theoretical predictions, the alloys of interest have been cast and characterized with respect to their structure, crystallographic texture, and elastic stiffness. Since the ab-initio simulations are applied to elucidate basic tendencies of suited alloy compositions under the constraints described on the one hand and to optimize the theoretical framework for such tasks on the other they are confined in a first step to a set of binary systems of Ti with different high melting bcc refractory transition metals, namely, Nb and Mo. The experiments are conducted both, on some of the binary alloys and also on two more complex engineering alloy variants, namely, Ti-35wt.%Nb-7wt.%Zr-5wt.%Ta and a Ti-20wt.%Mo-7wt.%Zr-5wt.%Ta.

The reasoning behind the use of ab initio calculations for understanding and designing novel β-Ti alloys pursues the following strategy: First, we hope that a more systematic theoretical inspection of the structure and thermodynamics of such alloys provides basic insight into the electronic tendencies as a function of compositions which provide a more detailed guideline to future alloy design approaches than commonly applied in the field of metallurgical alloy design. Second, we aim at replacing empirical rules for alloy design by rules which are built on the nature of the electronic bond. Third, we hope to reduce the huge efforts commonly associated with experimental alloy screening in that field (melting, casting, heat treatment, homogenization, thermodynamic and mechanical characterization) by using theoretical predictions of structure and elastic properties. Fourth, a better understanding of alloy design on the basis of ab initio simulations can reveal shortcuts to arriving at advanced promising alloy concepts as opposed to conventional metallurgical try and error methods. Fifth, it is our aim to develop and critically evaluate the applicability of ab initio predictions in the field of metallurgy by comparing the simulation results with experiments. Sixth, we want to proof that ab initio methods can not only be of great service to investigate the stability of certain single crystalline structures but that they can also predict mechanical properties which are of immediate engineering relevance for the final product design. Seventh, we want to show that in some cases – like in the present case of the design of β-Ti alloys – a scale jumping modeling strategy which directly predicts intrinsic properties such as structure and elasticity without the use of mesoscale methods, can for certain technological problems be more rewarding and efficient than multiscale modeling approaches which aim to combine predictions from many additional intermediate scales for the same aim.

## 2 Theoretical and Experimental Methods

### 2.1 Theoretical approaches

In order to identify Ti-based alloys which obey the constraints outlined in section 1 (i.e. non-toxic elements, stable in the β-phase, low Young modulus [1-30,34]) we have performed a systematic screening of binary alloys employing parameter-free density-functional theory calculations [31,32].

In this study, we restricted our research on Ti-Mo and Ti-Nb binaries with both Mo and Nb being non-toxic alloy ingredients. The aim of the calculations was to predict and understand metallurgical trends. In particular we were interested in a detailed understanding how the alloy composition affects the stability of the β-phase and whether and how it can be used to tailor the Young modulus. The calculations have been performed using a plane wave pseudopotential approach as implemented in the Vienna Ab-initio Simulation Package (VASP) code [35,36]. The plane wave cutoff energy has been 170 eV and a 8×8×8 Monkhorst-Pack mesh has been used to sample the Brillouin zone. The binary alloys have been described by supercells consisting of 2×2×2 elementary cubic or hexagonal unit cells with a total of 16 atoms. A large variety of alloy compositions has been studied by systematically replacing Ti atoms by either Nb or Mo atoms. The lowest alloy composition was 6.25at.% (one Nb/Mo atom in a 16 atom supercell). For each alloy composition various local arrangements have been considered and in total 48 body-centered cubic (bcc) and 28 hexagonal closed-packed (hcp) configurations have been studied.

For each of the constructed supercells the equilibrium geometry has been calculated, i.e., the geometry for which the total energy reaches a minimum and the forces on the atoms disappear and the system is strain free. Initially, we performed a full relaxation both, for the atoms (to bring the forces acting on the atoms to zero) and for the size and shape of the supercell (to make the system strain free). The biggest contribution turned out to be due to the volume relaxation of the cubic supercell. Relaxation effects which are due to changes in the cell shape (non-uniaxial components, shear components) or due to displacements in the atomic positions have been found to be small with respect to the energy differences due to the variations in the local arrangement. For the present study we have, therefore, neglected these relaxation effects and have taken into account only volume relaxation. The cell shape for the hexagonal hcp phase has been fixed by keeping the c/a ratio constant to the value of pure hcp Ti (c/a=1.59).

The fundamental quantity expressing the thermodynamic stability of an alloy is the alloy formation energy. The formation energy (per atom) of a Ti-binary alloy $Ti_xX_{1-x}$ (X=Nb,Mo here) in configuration σ = bcc/hcp is defined as:

(1)     $E^{\sigma}_f(Ti_xX_{1-x}) = E^{\sigma}_{tot}(Ti_xX_{1-x})/N - x \cdot \mu^{Ti}(\text{hcp-bulk}) - (1-x) \cdot \mu^{X}(\text{bcc-bulk})$.

Here, $N$ is the total number of atoms per supercell, $E^{\sigma}_{tot}(Ti_xX_{1-x})$ the total energy per supercell of the alloy, $\mu^{Ti/X}$ the chemical potential of element Ti or X in its corresponding bulk phase, respectively, and $x$ the alloy composition. In the above definition (Eq. (1)) the alloy is thermodynamically stable for $E_f < 0$.

In order to determine thermodynamic properties and stability of phases at elevated temperatures, as e.g. during the thermal treatment of the samples, the entropy effects have to be considered. These can be decomposed for solids into a configurational (mixing) and vibrational contribution [37]. In this study we will restrict on a rough estimate of the temperature dependence and neglect the vibrational contribution which is computationally difficult to access. The remaining contribution, the configurational entropy, will be calculated in the ideal mixing approximation. This approximation becomes exact for alloys where the formation energy depends only on the concentration but not on the local atomic configuration. As will be shown below, the approximation is well justified in case of the Ti-alloys studied here. The ideal mixing entropy is given by:

(2)     $S_{config}(x) = k_B \cdot [x \cdot \ln(x) + (1-x) \cdot \ln(1-x)]$

where $x$ is the alloy composition of a binary alloy $Ti_xX_{1-x}$ and $k_B$ the Boltzmann constant. The temperature dependent free energy can than be calculated according to

(3)     $F_f(x,T) = \langle E_f(Ti_xX_{1-x}) \rangle - T \cdot S_{config}$

where the averaged formation energy $\langle E_f(Ti_xX_{1-x}) \rangle$ is obtained from the formation energies of alloys with different local atomic configuration but same concentration by averaging using the Boltzmann statistics at the reference temperature. The temperature dependent free energy of formation allows to determine the thermodynamic stability of an alloy at a given temperature.

## 2.2 Experimental methods

The various Ti alloys were melted in an electric arc furnace. All of the used alloy elements had a high purity (table 1). The electric arc furnace was evacuated and subsequently flooded with Argon at a pressure of 300 mbar. The furnace was equipped with a water cooled copper crucible. The temperature of the electric arc amounted to about 3000°C while the melt was at the center hold at a peak temperature of 1830-1850°C in order to assure complete dissolution of the Nb, Ta, or Mo respectively. The electric arc method provided an intense stirring effect. Melting all ingredients required about 30-60s.

In order to obtain cast samples of optimal chemical and structural homogeneity all specimens were remelted several times. This means that each sample was after melting and stirring completely solidified in the crucible, then turned about its cross axis by use of an in-furnace manipulator, and subsequently reheated above the melting point. This sequential procedure (melting, stirring, solidification, rotation) was repeated four times. After the forth remelting step the sample was finally cast into a rectangular copper mold which had a size of 60mm×32,6mm×10mm. The copper mold had a temperature of about 30°C which led to rapid solidification entailing only microsegregation and suppressing dendrite formation.

Table 1. Alloy compositions

| | Ti20Mo7Zr5Ta (engin. notation: wt.%) | | Ti35Nb7Zr5Ta (engin. notation: wt.%) | | Ti10Nb (phys. notation: at.%) | | Ti20Nb (phys. notation: at.%) | | Ti25Nb (phys. notation: at.%) | | Ti30Nb (phys. notation: at.%) | | Ti10Mo (phys. notation: at.%) | | Ti20Mo (phys. notation: at.%) | |
|---|---|---|---|---|---|---|---|---|---|---|---|---|---|---|---|---|
| **Element** | wt% | at% | wt% | at% | wt% | at% | wt% | at% | wt% | at% | wt% | at% | wt% | at% | wt% | at% |
| Ti | 68 | 82,0 | 53 | 69.7 | 82.3 | 90 | 63.3 | 80 | 60.07 | 75 | 54.6 | 70 | 81.8 | 90 | 66.6 | 80 |
| Mo | 20 | 12,0 | | | | | | | | | | | 18.2 | 10 | 33.4 | 20 |
| Zr | 7 | 4,4 | 7 | 4,8 | | | | | | | | | | | | |
| Ta | 5 | 1,6 | 5 | 1,7 | | | | | | | | | | | | |
| Nb | | | 35 | 23,7 | 17.7 | 10 | 32.7 | 20 | 39.3 | 25 | 45.4 | 30 | | | | |

All as-cast samples were subjected to a solution heat treatment at 1473K (1200°C) for 3h in order to homogenize the sample and remove microsegregation. Since Ti alloys undergo very strong chemical reactions with oxygen the samples were generally heat treated under Argon atmosphere. The more complex engineering alloy Ti-20wt.%Mo-7wt.%Zr-5wt.%Ta was also solution annealed for 3h while the alloy Ti-35 wt.%Nb-7wt.%Zr-5wt.%Ta was solution annealed for 4h.

The ensuing characterization of the chemical and microstructural homogeneity of the cast and heat treated samples was conducted by using optical and scanning electron microscopy (SEM) in conjunction with EDX (energy dispersive x-ray spectrometry) and EBSD (electron back scatter diffraction) and x-ray Bragg diffraction methods.

Grinding before microscopy was performed using 400, 600, 1200, and 2400 paper with subsequent use of 6μm (45 minutes), 3μm (20 minutes), and 1μm (15 minutes) polishing. For optical microscopy the samples were etched using a solution of 68 ml Glycerin, 16 ml $HNO_3$, and 16 ml Hf. For SEM inspection the specimens were after the final 1μm polishing step finished by electrolytic polishing at 35V for 60s.

The elastic properties were investigated by using an ultrasonic resonance frequency method (Grindo-Sonic). This method measures the elastic modulus by analyzing the natural period of the transient vibration which results from a mechanical disturbance of the object tested. The Grindo-Sonic device transforms the incoming signal received from this natural frequency in an electric current of the same frequency and relative amplitude, during eight periods, due to a quartz clock where a reference crystal oscillates at a given frequency.

In order to determine the volume fractions of the two phases after thermal homogenization experimentally, we conducted x-ray wide angle diffraction experiments on the various alloys. The measurements were conducted either with $Mo_{K\alpha1}$ or $Co_{K\alpha1}$ radiation obtained from tubes operated at 40mA and 40kV. The crystallographic textures of the cast and homogenized samples were weak so that only a small corresponding error has to be taken into account.

## 3 Results and discussion

### 3.1 Thermodynamic analysis of phase stability: ab initio simulations

Based on Eq. (1), the T=0K formation energies for the various ordered α and β structures have been calculated (Figs. 1a and b). In order to analyze these data we first focus on the *relative stability* between the two phases. According to Figs. 1a and b both, Nb and Mo destabilize the α-phase (the formation energy increases with the composition $x$) but stabilize the β-phase (the formation energy decreases with composition $x$). Only for low concentrations of these elements (<10at.% for Mo, <20at.% for Nb) the α-phase is found to be energetically more favorable than the β-phase.

Let us now focus on the *thermodynamic stability*. For an alloy to be thermodynamically stable its formation energy must be negative. According to Figs. 1a and b Nb and Mo exhibit a qualitatively different behavior. For Ti-Nb alloys the formation energies at temperature T=0K of both the α and β phase are positive (endothermic) for Nb concentrations up to 93at.% (Figs. 1a). An important implication of this result is that at low temperatures this alloy is thermodynamically completely immiscible in this concentration regime. Only for compositions above 93at.% the β phase is intrinsically stable, i.e. the alloy formation energy is negative (exothermic) even without any entropy stabilization effects.

In contrast to the Ti-Nb system, for Ti-Mo alloys the T=0K formation energy is endothermic only for low Mo concentrations but becomes exothermic at Mo concentrations larger than $x_{crit} \approx 25$at.%. We, therefore, conclude that Mo acts above a critical concentration $x_{crit}$ as an *intrinsic* bcc stabilizer, i.e., it promotes the formation of the bcc phase even in the absence of any entropy driven temperature effects (Fig. 1a).

In order to estimate the effect temperature has on the formation / stabilization of the cubic phase the alloy formation energy has been calculated at finite temperatures employing Eq. (3). A crucial issue is to define a suitable theoretical reference temperature to be used in Eq. (3). This reference temperature can be regarded as limit: Below it the system can no longer maintain thermodynamic equilibrium and the phases which are thermodynamically stable at this temperature are frozen in rendering the system metastable. At first glance it might be appealing to use the experimental annealing temperature as a reference state. In the annealing process the samples are kept for three hours at this temperature which is sufficient for an alloy to reach thermodynamic equilibrium. However, the annealing temperature employed in our experiments (T=1200 °C) is well above the transition temperature of Ti (T=882 °C) and the thermodynamic ground state of Ti is no longer hcp but bcc. Therefore, we felt that a more suitable choice as reference is the hcp-bcc transition temperature of Ti. The corresponding results are shown in Figs. 1c and d.

According to Figs. 1a and b, the spread in energy for a given composition x but different local configurations is small. This observation indicates that for the binary alloys investigated the dependence of local configuration on the energy of formation is weak. Thus, the assumption underlying the approximation of ideal mixing as used to derive Eq. (3) is well justified.

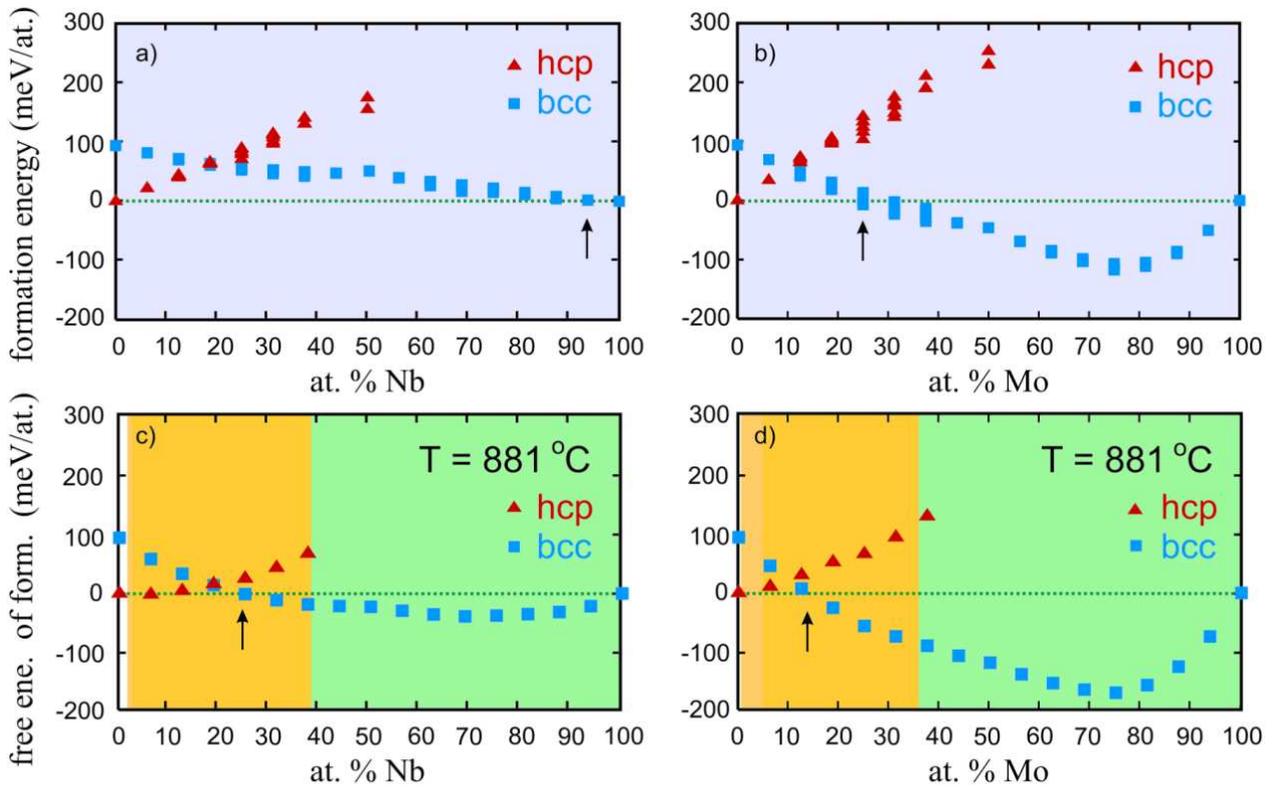

Figure 1. Theoretical alloy formation energies at T=0K (a-b) and free energies of formation at 1K below the bcc-hcp transformation temperature in pure Ti, T=881 °C (c-d) for binary Ti-X alloys as a function of the alloy composition. The images on the left hand side show the results for Ti-Nb and those on the right hand side for Ti-Mo alloys. The (red) triangles/(blue) squares mark the formation energies for the hcp/bcc phase. The color of the background in c and d indicates the co-existing phases as calculated in section 3.2: yellow: phase coexistence of hcp and bcc; blue: single-phase bcc. The arrow shows the critical concentration $x$ above which the bcc-phase becomes thermodynamically stable.

A comparison of Figs. 1a and c shows that going from T=0K to finite temperatures significantly reduces the alloy formation energy: While Nb stabilizes at T=0K the β-phase only at Nb concentrations well above 93%, finite temperature effects strongly reduce the formation energy. As shown in Figs. 1c, close to the transformation temperature the free energy of formation of β-Ti-Nb alloys becomes negative (exothermic) even for Nb concentrations as small as 25at.%. We can, therefore, conclude that Nb acts as an *entropy-driven* β-stabilizer.

For Mo, which acts even at T=0K and in modest concentrations as a β-stabilizer (Fig. 1b) finite temperature effects are less pronounced (Fig 1d): The minimum Mo concentration at which the alloy formation energy becomes zero reduces from 25at.% to 14at.%.

The fact that Mo is a better β-stabilizer then Nb is in agreement with earlier experimental studies where Mo was found to stabilize the β-phase already for Mo concentrations exceeding 5at.% whereas 24at.% of Nb was necessary to achieve the same goal [39]. A direct quantitative comparison between our theoretical data and the experimental data in Ref. [39] is not straightforward. The determination of the first occurrence of a stable fraction of the β-phase by Dobromyslov and Elkin [39] was done by using an X-ray wide angle Bragg diffraction set-up in conjunction with Cu $K_{\alpha 1}$ radiation. This method is precise usually only to a value of 1-3vol.%. Also, the presence of crystallographic textures may usually further reduce the precision of the x-ray analysis for small amounts of a second phase. Also, Cu $K_{\alpha 1}$ radiation may create minor false intensity contributions such as caused by fluorescence.

**3.2 Thermodynamic analysis: comparison of the theoretical results and experimental data**

In order to address these issues we have experimentally not only determined the onset of the β-phase but also its volume fractions for given alloy composition. An advantage of this approach is that this quantity can be easily obtained from the calculated formation energies by a Gibbs construction.

To actually perform the Gibbs construction, the compositional dependence of the averaged formation energy (Figs. 1c, d) has been approximated by a third order polynomial using a least-square approach. The Gibbs constructions for both alloys and the computed volume fraction of the β-phase are summarized in Fig. 2. The results have been also used in Fig. 1c and d to determine the different phase regions (color-coding of the background). Employing the Gibbs construction we find that the threshold concentrations above which the alloys consist solely of the β-phase is ~39at% of Nb and ~36at% of Mo (Fig 2).

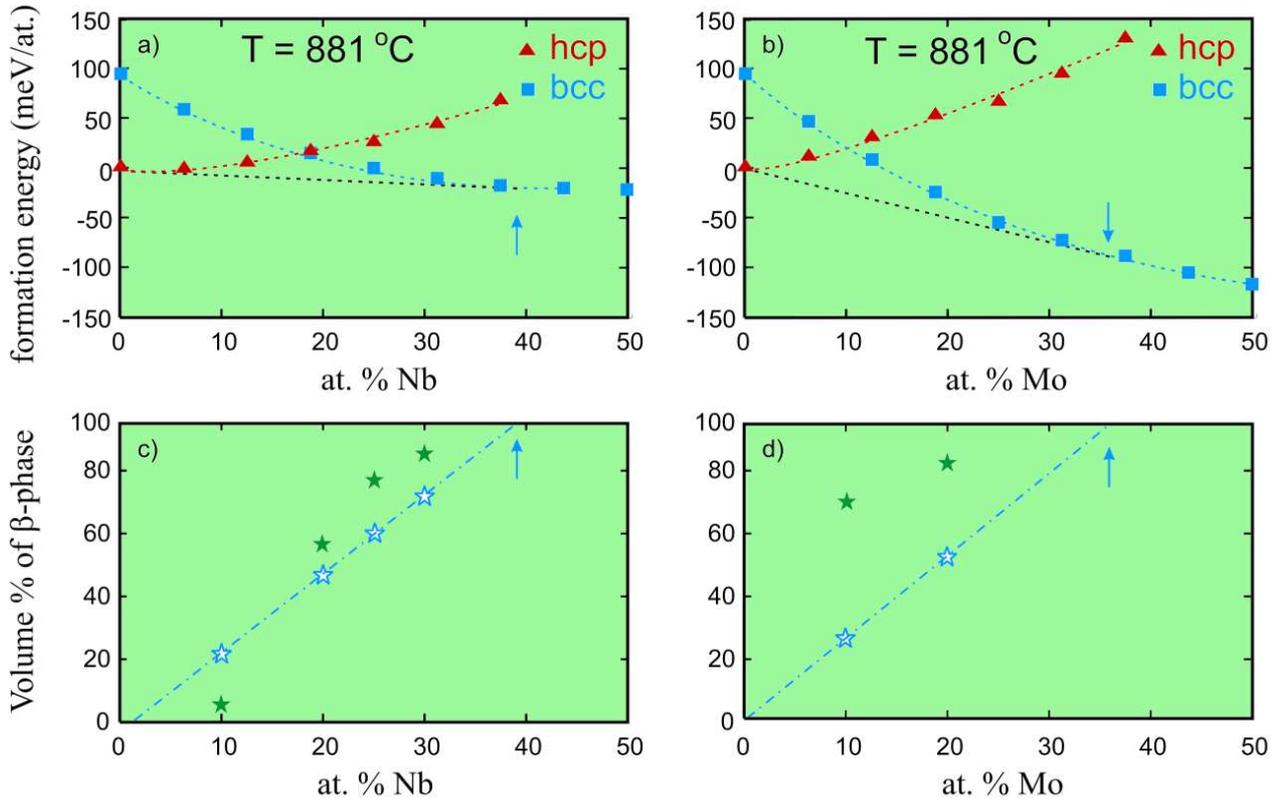

Figure 2
Gibbs construction (a,b) and volume fraction of the β-phase (c,d). Both binary alloy systems, Ti-Nb (a,c) and Ti-Mo (b,d) are shown. The volume fraction of the β-phase as determined by the ab initio method is shown by the (blue) dash-dotted line and empty blue stars for selected concentrations. The experimental data points (Table 2) are marked by full green stars. The blue arrow marks the threshold concentrations for single-phase bcc alloys. The error bars for the experimental data fall within the symbol size.

In order to experimentally determine the volume fractions of the two phases after thermal homogenization, we conducted x-ray wide angle diffraction experiments for the binary Ti-Nb and Ti-Mo alloys, Fig. 3. The results, which are summarized in Table 2, are shown as full green stars in Figs. 2(c,d). The experimental upper limits of the screened alloys are 88 vol.% β-phase for the Ti-20at.%Mo alloy and 90 vol.% β-phase for the Ti-30at.%Nb alloy.

For Ti-Nb alloys (Fig. 2c) the agreement between experiment and theoretical predictions is excellent. For Ti-Mo alloys, the agreement is less satisfactory. The discrepancy between theory and experiment may be due to (i) neglecting vibrational entropy which destabilizes the α-phase with increasing temperature, or (ii) the existence of another phase not been included in our analysis.

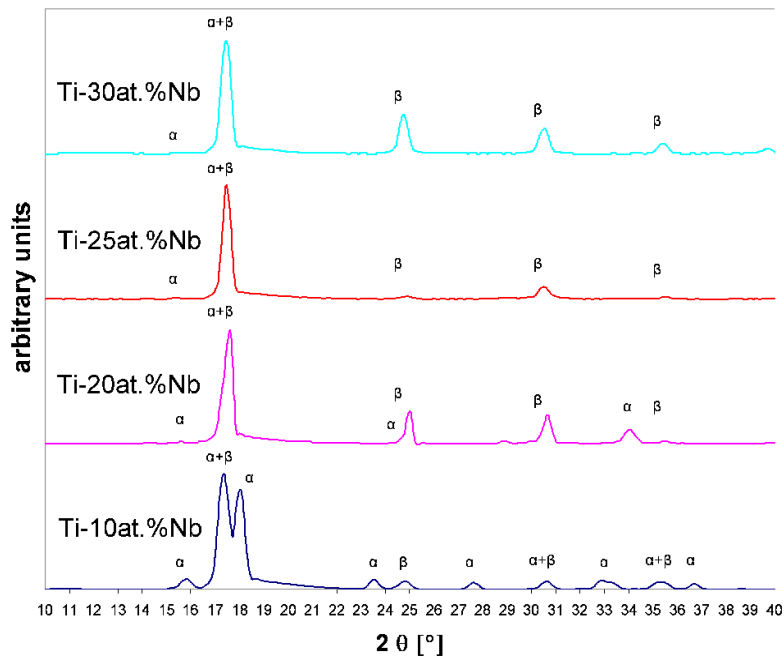

a)

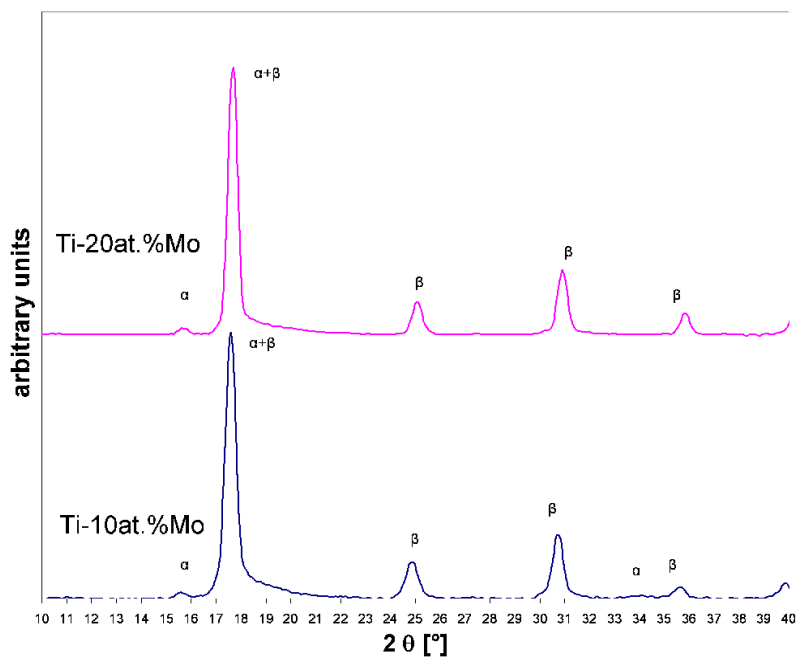

b)

Figure 3
Wide angle x-ray Bragg diffraction spectra for some of binary Ti-Nb and Ti-Mo alloys ($Mo_{K\alpha 1}$ radiation, 40mA, 40kV). The intensity is given in arbitrary units as a function of the diffraction angle $2\theta$.
a) binary Ti-Nb alloys
b) binary Ti-Mo alloys

Table 2. Experimentally observed volume fractions of the phases based on the current measurements with Mo K$_{\alpha 1}$ radiation

| Phase content by x-ray Bragg diffraction | Ti-10at.%Nb | Ti-20at.%Nb | Ti-25at.%Nb | Ti-30at.%Nb | Ti-10at.%Mo | Ti-20at.%Mo |
|---|---|---|---|---|---|---|
| volume fraction in %, α-phase (hcp) | 94 | 40 | 19 | 10 | 26 | 12 |
| volume fraction in %, β-phase (bcc) | 6 | 60 | 81 | 90 | 74 | 88 |

An interesting conclusion can be drawn by inspecting closer the Gibbs construction. According to Fig. 2b the Mo concentration in the α-phase should be close to zero. In order to verify this prediction, the phase content determined via x-ray diffraction was further cross-examined by joint EBSD and EDX measurements (see Fig. 4). A significant difference between the Mo content in the α and β phases (Mo depleted in the latter) was indeed observed.

The EBSD images presented in Fig. 5 give an example of the experimentally observed microstructures for one of the binary alloys, namely, for Ti-30at.% Nb. Fig 5a shows the as cast sample while Fig. 5b shows the sample after chemical homogenization for 3h at 1200°C. One has to note that the size bar is different for the 2 micrographs owing to grain growth that took place during homogenization (Fig 5b). The EBSD images reveal an isotropic grain shape and a random texture in either case (before and after homogenization). The color code is the Miller index as indicated in the standard triangle of lattice directions pointing in normal (i.e. casting) direction. The Kikuchi indexing procedure suggests that in the present case of the Ti-30at.%Nb alloy about 98% of all EBSD points revealed a bcc structure which is even larger than the value shown in Table 2 for the x-ray results. For samples with lower Nb content larger fractions of the α- and ω phase were found.

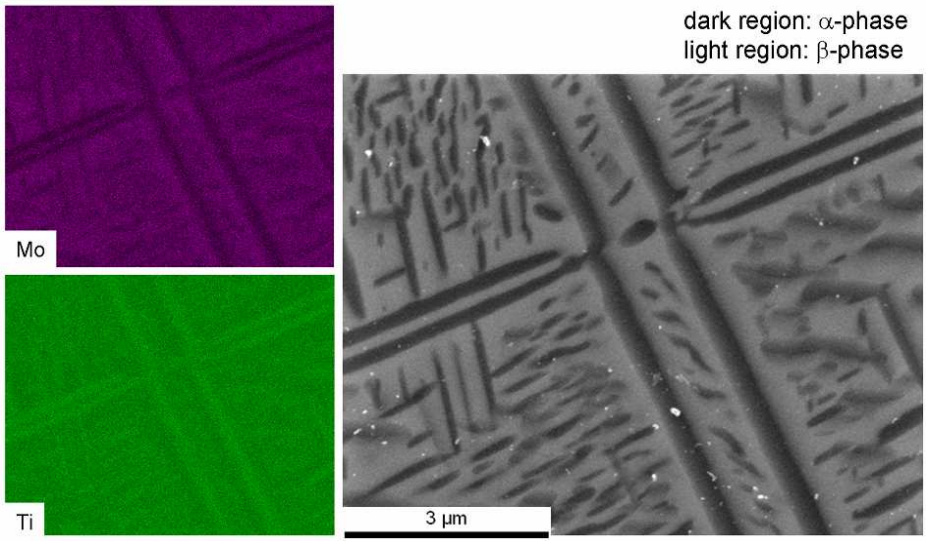

Figure 4
EDX measurements in the two-phase microstructure of the Ti-10%Mo sample which reveals close to zero intensity of Mo in the α-phase while the β-phase contains a high concentration of Mo. The sample was heat treated at 873 K for 4 days before the EDX measurement.

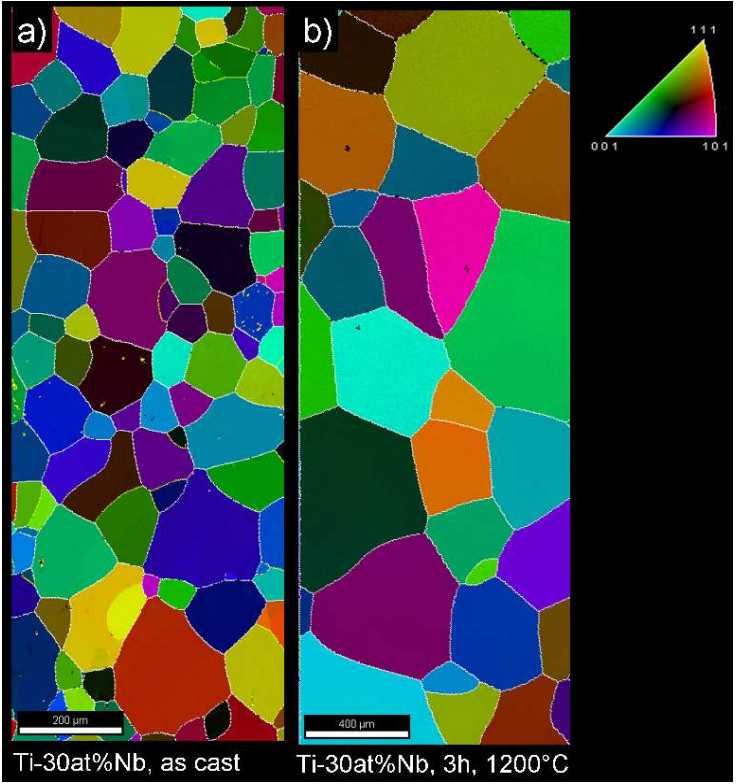

Figure 5
EBSD images as an example of the experimentally observed microstructures of one of the binary alloys, namely, of the Ti-30at.% Nb alloy. (a) shows the as cast sample; (b) shows the sample after homogenization for 3h at 1200°C. Note that the size bar is different for the 2 micrographs owing to grain growth that took place during homogenization. The EBSD images reveal an isotropic grain shape and a random texture. Color code: Miller index in standard triangle of lattice directions pointing in normal / casting direction.

**3.3 Ab initio predictions of the elastic properties in the binary systems Ti-Nb and Ti-Mo**

In order to identify metallurgical trends in the two binary alloy systems Ti-Nb and Ti-Mo with respect to the desired mechanical properties (reduced stiffness) the Young modulus in the crystallographic [001] direction has been calculated and analyzed for different binary alloy compositions. The theoretical predictions have been based on the simulation of an uniaxial tensile test [40-42] along the crystallographic [001] direction of the β-phase. To do so, we have proceeded in two steps. First, we determined the total energy of a given binary alloy in the ground state. In the second step, we applied an elongation along the [001] loading axis by a small fixed amount ε which is in linear approximation (Hooke law) equivalent to the application of an elastic tensile stress σ. For each value of the elongation ε the total energy was minimized by relaxing two stresses in the directions perpendicular to the loading axis. The stress σ in the loading direction [001] is given by

(4)    $\sigma = (\delta E/\delta \varepsilon)/(A \cdot a_0)$

where $E$ is the total energy per formula unit of the crystal, $a_0$ is the ground state lattice parameter and $A$ is the area (proportional to the formula unit of the crystal) perpendicular to the applied stress for each value of the elongation ε. The Young's modulus is then defined as the ratio between the stress σ and elongation ε in the limit of a zero strain load. Applying this approach, the Young's modules along the soft [001] direction have been calculated for both binary alloys (Fig. 6; blue squares). Also included in Fig. 6 are the experimental data points obtained from our ultrasonic measurements (green stars).

The simulated data in Fig. 6 reveal that the Young's modulus of the bcc Ti phase (β-phase) increases in both cases almost linearly with the alloy composition. The scatter of the predicted modulus data stems from the variation of the atomic configuration in the cell which was in all cases varied in order to evaluate the dependence of the results on short range order effects. It is interesting to note that for the lowest alloy compositions (i.e., close to the limit of pure Ti) the Young modulus for the bcc phase becomes negative. A negative Young modulus implies that for low Nb (<8at.%) or Mo (<4at.%) concentrations these bcc alloys are mechanically unstable against tetragonal elastic loadings. This conclusion is in agreement with previous studies on the elastic stability of bcc Ti (e.g. [43]) and reflects the fact that weakly alloyed Ti-Nb and Ti-Mo materials have the tendency to switch back into the stable hexagonal phase when forced in the simulation into a bcc structure, Fig. 1. The data points for the bulk polycrystal elastic data shown in Fig. 6 in comparison to the predicted [001] elastic stiffness data have been taken from samples which have before been exposed to the 3h, 1200°C homogenization treatment. These long heat treatments have been made in order to approximate an thermodynamic equilibrium state in the samples and, thereby, to achieve better comparability with the theoretical predictions. In all cases, the experimentally determined Young's modulus is slightly larger than the theoretically estimated results. The origin for this discrepancy is that the theoretical studies had been restricted to bulk single crystals while the real samples are polycrystalline (see Fig. 5). In a polycrystalline sample with random crystallographic texture the Young's module is isotropic and given by averaging the Young's moduli over a huge number of grains along all crystal orientations. The fact, that averaging includes soft and hard directions, whereas the theoretical study has focused on a single soft direction at T=0K, may explain the discrepancy between theory and experiment.

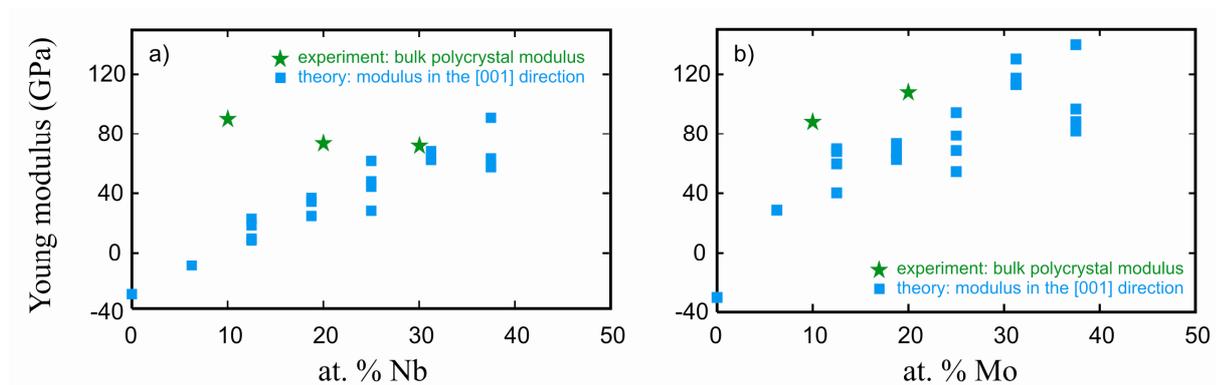

Figure 6
Predicted and experimentally obtained Young's moduli for two binary Ti alloy systems as a function of the alloy composition. The left figure (a) shows the results for Ti-Nb and the right one (b) for Ti-Mo alloys. The blue squares show the simulation results for the soft [001] crystal direction of the bcc lattice cell. The green stars in both images represent experimental results obtained on our polycrystalline samples by ultrasonic measurements. The error bars fall within the symbol size. The experimental elastic modulus data were taken on samples which were cast and subsequently 3h homogenized at 1200°C.

In the Ti-Mo system the experimentally observed moduli for the new binary randomly oriented polycrystalline alloys amount to 106.9 GPa for Ti-20at.%Mo and 88.4 GPa for Ti-10at.%Mo. In the Ti-Nb system the measured moduli are 72.1 GPa for Ti-30at.%Nb, 75.8 GPa for Ti-20at.%Nb, 91.2 GPa for Ti-10at.%Nb, i.e. the value observed for the Ti-30at.%Nb alloy is the smallest one of all alloys which were experimentally inspected. It must be noted that those materials with a small alloy content (10at.%) do most likely also contain substantial contributions of the α and ω-phases, i.e. the stiffness data do not reflect the values of a pure bcc structure. In comparison, the bulk polycrystal modulus of a hexagonal polycrystalline Ti sample which we measured as a reference amounted to 114.7 GPa. This means that the best of the new binary materials, namely Ti-30at.%Nb, yields a drop in stiffness by about 37%.

## 3.4 Design of more complex engineering alloys on the basis of the theoretical predictions

*Basic Aspects*

According to our novel strategy of a faster and more efficient way for bottom-up alloy design as outlined in section 1.2 we have in the second step of this work used the ab initio predictions obtained on the binary materials (Ti-Mo, Ti-Nb) as a starting point for the synthesis of two more complex engineering alloy variants, namely, Ti-35wt.%Nb-7wt.%Zr-5wt.%Ta and Ti-20wt.%Mo-7wt.%Zr-5wt.%Ta.

To be more specific in this discussion it makes sense to properly identify those aspects that were learned from the ab-initio simulations conducted on the two binary alloy systems with respect to more complex alloys design.

The first point clearly demonstrated by the simulations are the minimum threshold concentrations of the occurrence of a stable β-phase (together with a dominant α-phase) and also of the stable single-phase existence of the β-phase. The first aspect is important when aiming at the design of a dual phase alloy and the second one counts when aiming at a pure β-phase alloy with the smallest possible elastic stiffness (because of the bcc structure). In that context it was clearly observed from the simulations (matching experimental observations from the literature and our own measurements) that Mo has a stronger effect on the formation of the β-phase than Nb. In other words Mo is a more efficient bcc stabilizer in the Ti matrix than Nb. Another point pertaining to this aspect is the fact Ti-Mo bcc alloys are above a critical concentration of about 25at.% *intrinsically* stable without the aid of the mixing entropy while Ti-Nb alloys are in general only entropy stabilized. This means that the simulations clearly indicate the different thermodynamic quality for the two different bcc stabilizers.

This observation from the simulations is of great practical importance for engineering alloy design since Ti alloys are typically in a strong non-equilibrium state when cast and, hence, require careful homogenization treatments. This means that a more fundamental knowledge on the stability of different alloy variants helps a lot to properly design adequate heat treatments for alloy homogenization. Similar aspects apply to diffusion kinetics.

Another essential though more practical aspect of the thermodynamic insight into the stability of the alloys lies in choosing those alloying elements which stabilize the β-phase at the smallest possible alloy content, costs, and / or density increase.

The second group of observations we used from the simulations to design more complex materials along the lines prescribed by the two binary alloys is the evolution of the elastic modulus as a function of the alloying content. The advantage of predicting not only thermodynamic but also mechanical data associated with our theory-guided alloy design approach takes us beyond earlier conventional thermodynamic approaches used before for alloy design such as offered by Calphad and similar methods. The ab initio predictions of the elastic modulus clearly showed that the Ti-Nb bcc structure is less stiff than the Ti-Mo bcc structure.

This result reveals a conflict: On the one hand the low elastic modulus provided by a Ti-Nb-based bcc structure is highly desired for biomedical applications but on the other hand we found that this system is thermodynamically less stable. For more complex alloy design we can, hence, learn that a Ti-Nb-based material should contain some additional alloying content of another bcc stabilizer such as Ta, Mo, or W. This strategy would most likely provide a thermodynamically more stable bcc alloy with a small elastic stiffness.

Another essential feature associated with the bottom up-screening of the elastic properties is the possibility of predicting and designing also the elastic anisotropy. Although this aspect will be the subject of another publication the possibility for such a procedure is obviously at hand. First tests (which are not included in this work) have revealed that not only the absolute values of the elastic properties but also the anisotropy strongly changes as a function of the alloy content. This point can hence be exploited for further alloy design.

A further point that can be learned form the binary alloy predictions with respect to more complex Ti-alloy design is the fact that the desire for a low elastic modulus should be combined with high strength and hardness. These constraints usually are contradictory since the resistance to dislocation motion and dislocation multiplication (strength and hardness) scales linearly with the magnitude of the shear modulus (in the isotropic limit of linear elasticity). A possibility to escape from this difficulty lies in a solid solution hardening strategy. This effect is in any case already provided by the Nb or, respectively, Mo content required for the basic stabilization of the bcc phase, and, for the case of more complex alloys, by minor addition(s) of a second (or even second and third) bcc stabilizer (see argumentation above) such as Ta and / or W in solid solution. Another element which could serve for solid solution hardening would be Zr. Further alloy alternatives are quite limited by the constraint of optimum biocompatibility.

*Application to Engineering Alloy Design*

Using the ab initio results in the spirit discussed above together with corresponding further alloy suggestions from the literature [3-30] we designed, cast, processed, and characterized the two engineering alloy variants Ti-35wt.%Nb-7wt.%Zr-5wt.%Ta and a Ti-20wt.%Mo-7wt.%Zr-5wt.%Ta. The EBSD maps in Fig. 7, taken for 3 perpendicular sections, reveal that the texture of the cast material is practically random.

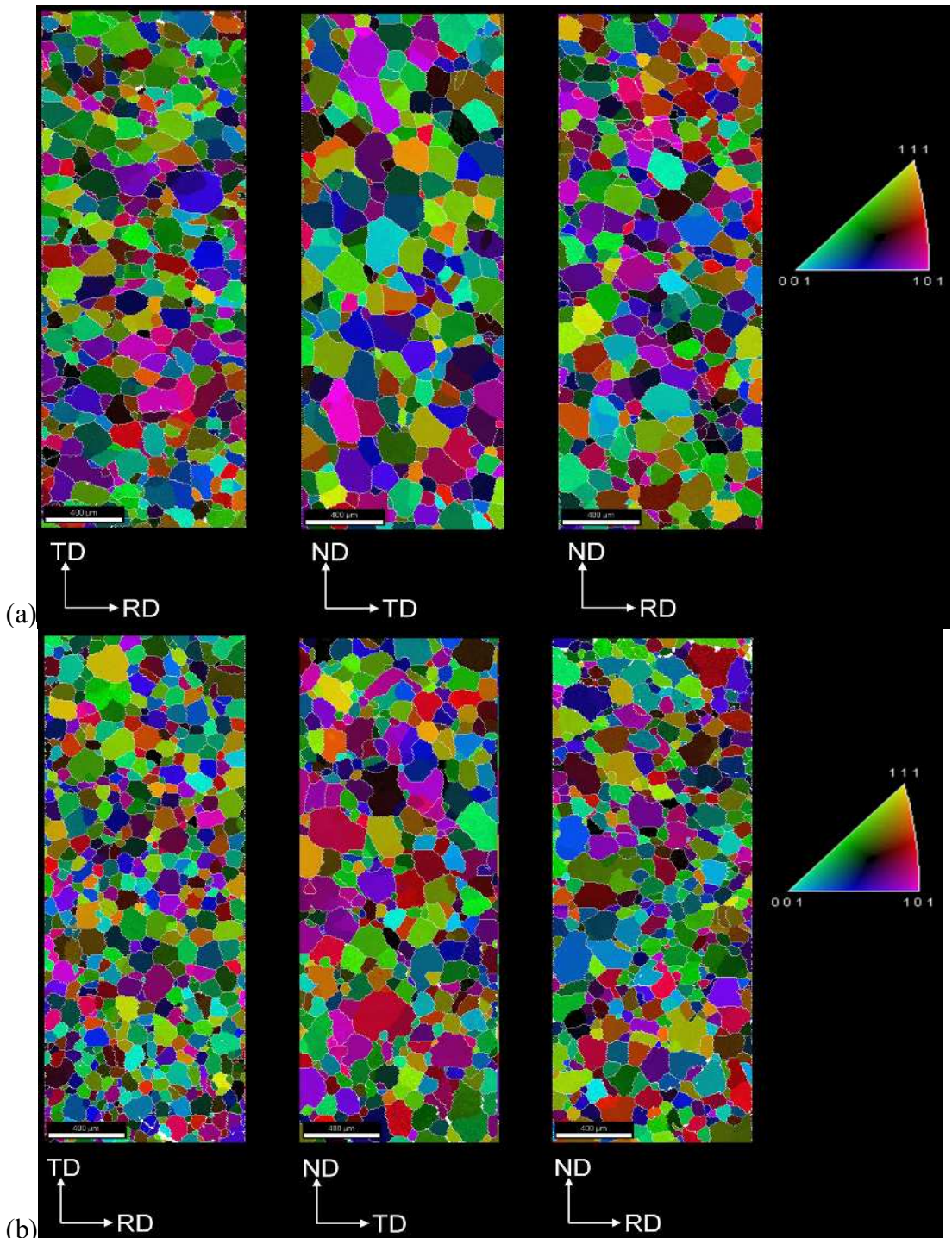

Figure 7
(a) EBSD images (normal plane indexed, data taken from 3 perpendicular directions) as an example of the experimentally observed microstructure of the as cast alloy Ti-35wt.%Nb-7wt.%Zr-5wt.%Ta.
(b) EBSD images (normal plane indexed, data taken from 3 perpendicular directions) as an example of the experimentally observed microstructure of the as cast alloy Ti-20wt.%Mo-7wt.%Zr-5wt.%Ta.
ND: normal direction; RD: longitudinal direction; TD: transverse direction.

Both specimens revealed pronounced microsegregation in the as cast state which had to be removed by corresponding homogenization heat treatments at 1473K (1200°C) under Argon atmosphere. The alloy Ti20wt.%Mo7wt.%Zr5wt.%Ta was solution annealed for 3h and the alloy Ti35wt.%Nb7 wt.%Zr5 wt.%Ta was solution annealed for 4h, Fig. 8.

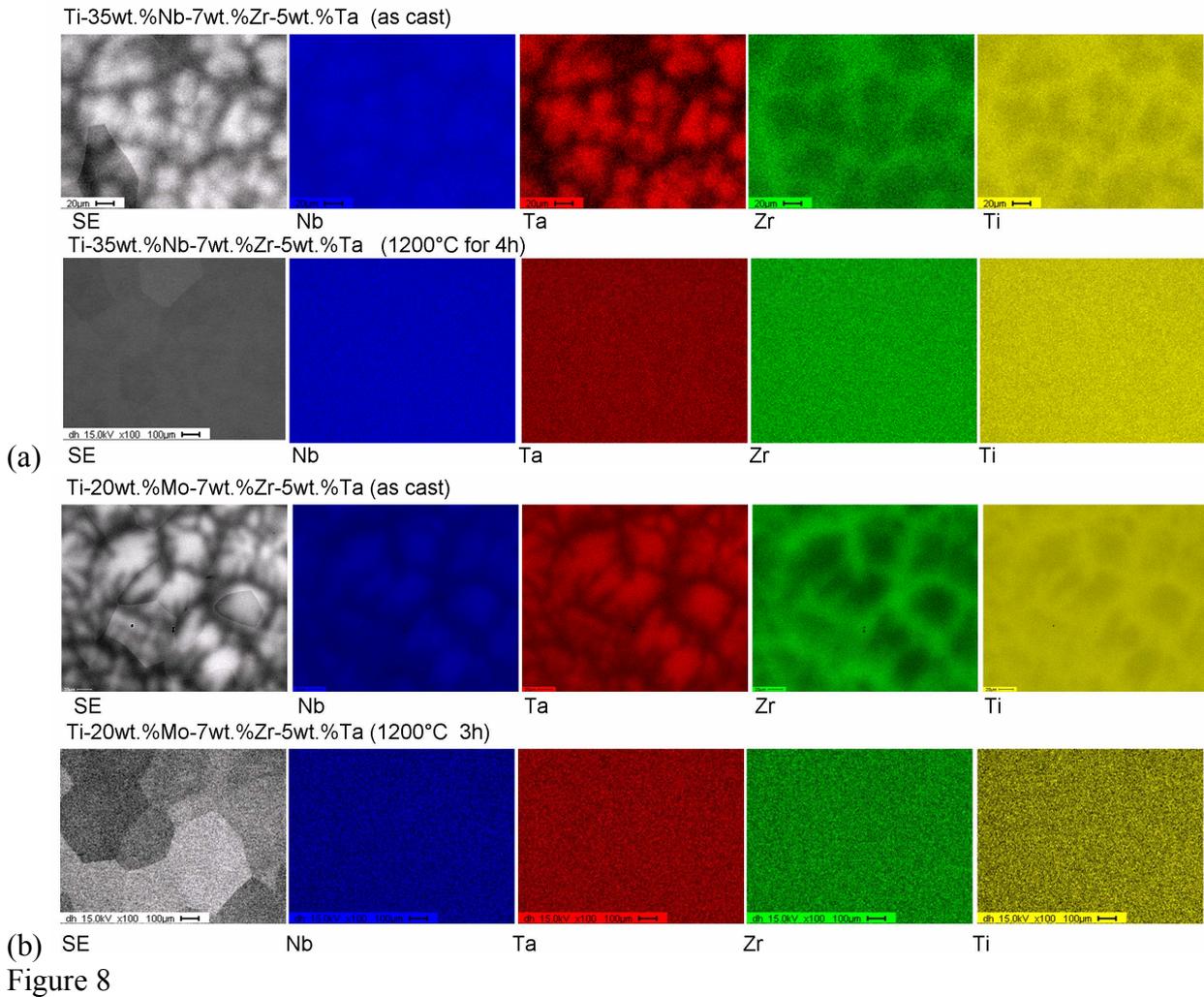

Figure 8
(a) EDX data before and after the heat treatment for the alloy Ti-35wt.%Nb-7wt.%Zr-5wt.%Ta.
(b) EDX data before and after the heat treatment for the alloy Ti-20wt.%Mo-7wt.%Zr-5wt.%Ta.

The x-ray spectra after heat treatment showed that the material consisted essentially of the bcc phase. The volume fractions of the α- and ω phases very small in both cases, Fig. 9.

While the bulk polycrystal modulus of the pure hexagonal Ti reference sample amounted to 114.7 GPa that of the heat treated Ti20wt.%Mo7wt.%Zr5wt.%Ta alloy was 81.5 GPa and that of the heat treated Ti35wt.%Nb7 wt.%Zr5 wt.%Ta alloy as low as 59.9 GPa.

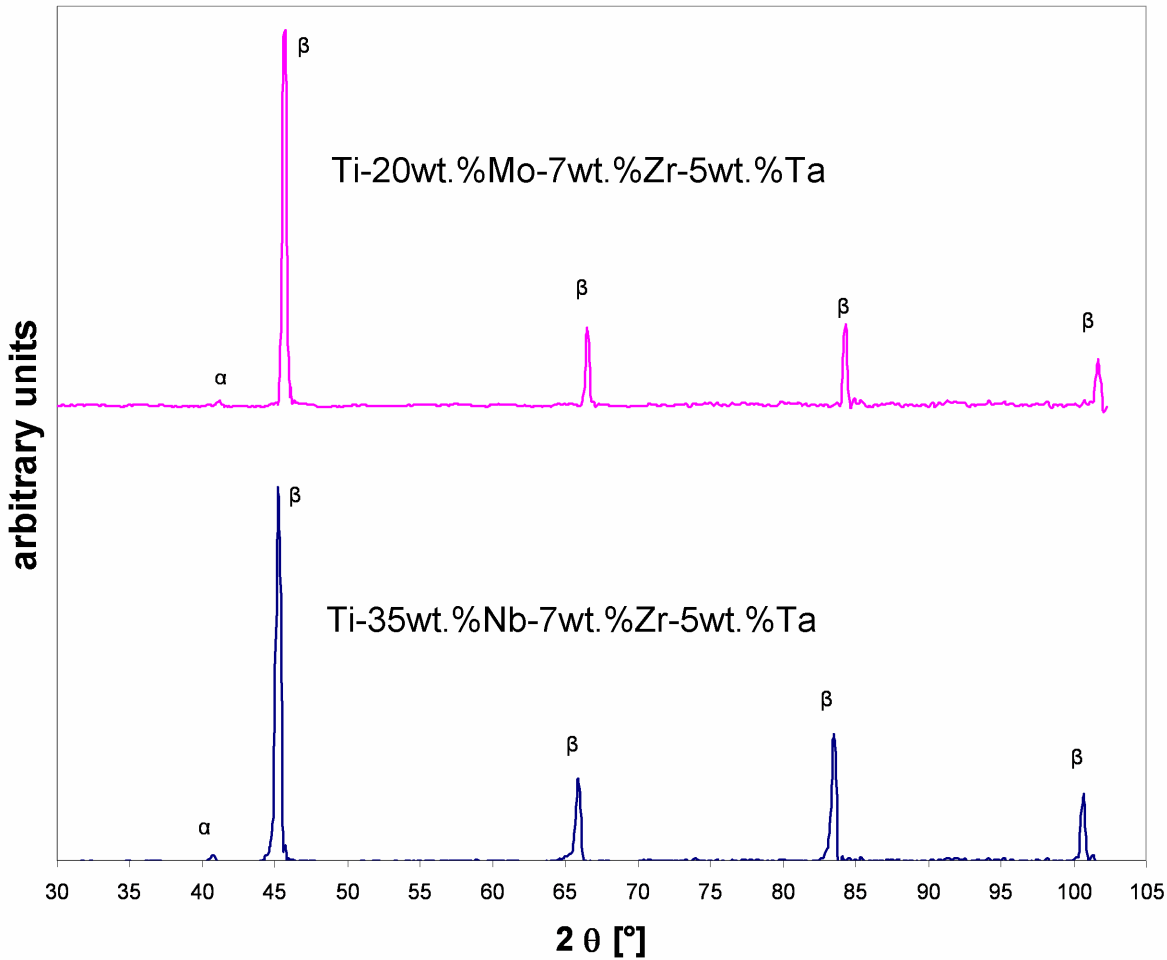

Figure 9
Wide angle x-ray Bragg diffraction spectra of the two engineering alloys Ti-20wt.%Mo-7wt.%Zr-5wt.%Ta and Ti-35wt.%Nb-7wt.%Zr-5wt.%Ta after heat treatment (Co$_{K\alpha1}$ radiation, 40mA, 40kV). The intensity is given in arbitrary units as function of the diffraction angle 2 θ.

### 3.5 Discussion of the suitability of ab initio based design strategies in physical metallurgy

Both, the experimental results obtained for the binary and for the engineering quaternary alloys clearly justify the use of ab initio simulations of the ground state energy and of the temperature-dependent free energy for modern materials alloy design [44-49]. In both cases (Ti-Nb, Ti-Mo) the basic tendency of the thermodynamic stability and of the influence of the entropy have been properly predicted. Not only our own current experimental results but also previous experimental observations reported in the literature show that the tendency of Nb and Mo to form a stable bcc Ti alloy at higher alloy content are correct. For the Ti-Nb binary alloy system it was also found, that although the Nb has the advantage of a strong reduction of the elastic modulus, it is not an intrinsic stabilizer of the β-phase but requires the contribution of the mixing entropy. According to the ab initio predictions for the Ti-Mo alloy system the β-phase becomes even intrinsically stable for Mo concentrations in the range around 25at.% without the contribution of the mixing entropy. This means that, although the basic tendency of high melting bcc refractory metals to stabilize the bcc phase of Ti was already observed before in the literature on the basis of empirical experience, the current ab initio simulations have the clear advantage to much better reveal the true stability of the bcc phase created.

An aspect which might deserve deeper future analysis in that context of the free energy of Ti bcc phase is the possible role also of the vibrational entropy contribution for the stabilization of the β-phase which may be relevant in the calculation of the free energy as known from some other bcc materials. In the current study we skipped that contribution owing to the very long and complex simulation procedures which are required to cover this point in depth. It will be addressed, though, in a separate study.

Beyond this clearly proven basic thermodynamic advantage of the new method for modern alloy design also the prediction of the elastic stiffness by the ab initio tensile test is of great relevance. It was shown that firstly, it is in principle possible to screen the elastic properties of new alloys before actually melting and casting them, and secondly, that the simulations are also in good agreement with the experimental observations.

# 4    Conclusions

The study presented a new strategy for the theory-guided ab initio based bottom up design of β-Ti alloys for biomedical applications using a quantum mechanical approach in conjunction with experiments. The method was applied to a set of binary Ti-Nb and Ti-Mo alloys. Following the predictions, the alloys were cast and characterized with respect to their crystallographic structure, microstructure, texture, and elastic stiffness. In addition we also studied two related engineering alloy compositions, namely, Ti-35wt.%Nb-7wt.%Zr-5wt.%Ta and a Ti-20wt.%Mo-7wt.%Zr-5wt.%Ta. The results show that ab initio simulations are very well suited for the bottom up design of new alloys. Both the thermodynamic stability and the elastic properties of the binary alloys could be well predicted using ground state and free energy calculations on the one hand and an ab initio elastic tensile test on the other.